\documentclass[aps,prb,showpacs,twocolumn,superscriptaddress]{revtex4}
\usepackage{amsmath,graphicx,epsfig,amssymb,subfigure,times,dsfont,algorithmic}
\usepackage[usenames]{color}

\newcommand{\ket}[1] {| #1 \rangle}
\newcommand{\bra}[1] {\langle #1 |}

\newcommand {\Fig}[1] {Fig.~\ref{#1}}
\newcommand {\Eqn}[1] {Eq.~(\ref{#1})}
\newcommand {\Sec}[1] {Sec.~\ref{#1}}

\begin{document}

\title{Variational Monte Carlo with the Multi-Scale Entanglement Renormalization Ansatz}

\author{Andrew~J.~Ferris}
\affiliation{The University of Queensland, School of Mathematics and Physics, Queensland 4072, Australia}
\affiliation{D\'epartement de Physique, Universit\'e de Sherbrooke, Qu\'ebec, J1K 2R1, Canada}
\author{Guifre~Vidal}
\affiliation{The University of Queensland, School of Mathematics and Physics, Queensland 4072, Australia}
\affiliation{Perimeter Institute for Theoretical Physics, Waterloo, Ontario, N2L2Y5, Canada}

\date{\today}
\begin{abstract}
Monte Carlo sampling techniques have been proposed as a strategy to reduce the computational cost of contractions in tensor network approaches to solving many-body systems. Here we put forward a variational Monte Carlo approach for the multi-scale entanglement renormalization ansatz (MERA), which is a unitary tensor network. Two major adjustments are required compared to previous proposals with non-unitary tensor networks. First, instead of sampling over configurations of the original lattice, made of $L$ sites, we sample over configurations of an effective lattice, which is made of just $\log (L)$ sites. Second, the optimization of unitary tensors must account for their unitary character while being robust to statistical noise, which we accomplish with a modified steepest descent method within the set of unitary tensors. We demonstrate the performance of the variational Monte Carlo MERA approach in the relatively simple context of a finite quantum spin chain at criticality, and discuss future, more challenging applications, including two dimensional systems.
\end{abstract}


\maketitle

\section{Introduction}

Understanding the collective behaviour of quantum many-body systems remains a central topic in modern physics, as well as one of the greatest computational challenges in science. Quantum Monte Carlo sampling techniques are capable of addressing a large class of (unfrustrated) bosonic and spin lattice models, but fail when applied to other models such as frustrated antiferromagnets and interacting fermions due to the so-called sign problem. Variational approaches, on the other hand, are sign-problem free but are typically strongly biased towards specific many-body wavefunctions. An important exception is given by the density matrix renormalization group,\cite{White1992} a variational approach based on the matrix product state (MPS),\cite{Oestlund1995a} which is capable of providing an extremely accurate approximation to the ground state of most one-dimensional lattice models. The success of DMRG is based on the fact that an MPS can reproduce the structure of entanglement common to most ground states of one-dimensional lattice models.

In order to extend the success of DMRG to other contexts, new tensor networks generalizing the MPS have been proposed. For instance, the multi-scale entanglement renormalization ansatz (MERA),\cite{Vidal2007b} with a network of tensors that extends in an additional direction corresponding to length scales, is particularly suited to address quantum critical systems.
Most significant has also been the proposal of tensor networks for systems in two and higher dimensions, where the MPS becomes inefficient. Scalable tensor networks include the projected entangled-pair states PEPS\cite{Verstraete2004} (a direct generalization of the MPS to larger dimensions) and higher dimensional versions of the MERA.\cite{Evenbly2009,Cincio2008} They can be used to address frustrated antiferromagnets and interacting fermions, since they are free of the sign problem experienced by quantum Monte Carlo approaches.

In a tensor network state, the size of the tensors is measured by the bond dimension $\chi$. This bond dimension $\chi$ indicates how many variational coefficients are used. Crucially, it also regulates both the cost of the simulation, which scales as $\mathcal{O}(\chi^p)$ for some large power $p$, and how much ground state entanglement the many-body ansatz can reproduce. In the large $\chi$ regime, PEPS and MERA are essentially unbiased methods, but with a huge computational cost that is often unaffordable. More affordable simulations are obtained in the small $\chi$ regime, but there these methods are biased in favour of weakly entangled phases (e.g. symmetry-breaking phases) and against strongly entangled phases (e.g. spin liquids and systems with a Fermi surface). Identifying more efficient strategies for tensor network contraction, so that larger values of the bond dimension $\chi$ can be used and the bias towards weakly entangled states is suppressed, is therefore a priority in this research area.

Refs. ~\onlinecite{Schuch2008}, \onlinecite{Sandvik2007} proposed the use of Monte Carlo sampling as a means to decrease computational costs in tensor network algorithms. [We note that there are other variational ans\"atze, such as so-called correlated product states, entangled plaquette states, and string-bond states, whose contractibility relies on sampling; see the introduction of Ref. \onlinecite{Changlani2009} for a review]. In a tensor network approach such as MPS, MERA or PEPS, sampling over specific configurations of the lattice allows to reduce the cost of contractions (for single samples) from $\mathcal{O}(\chi^p)$ to $\mathcal{O}(\chi^{q})$, where $q$ is significantly smaller than $p$, typically of the order of $p/2$. Needless to say, sampling introduces statistical errors. However, if less than $O(\chi^{q-p})$ samples are required in order to achieved some pre-established accuracy, then overall sampling results in a reduction of computational costs.

The proposal of Refs. ~\onlinecite{Schuch2008}, \onlinecite{Sandvik2007} is based on computing the overlap of the tensor network state with a product state (representing the sampled configuration). As such, it cannot be directly applied to the MERA, because the overlap of a MERA with a product state cannot be computed efficiently. Luckily, as discussed in Ref.~\onlinecite{Ferris2011a}, a sampling strategy specific to unitary tensor networks (such as MERA and unitary versions of MPS and tree tensor networks) is not only possible, but it actually has several advantages. Most notably, sampling takes place over configurations of a reduced, effective lattice; and it is possible to perform perfect sampling, by means of which uncorrelated configurations are drawn directly according to the correct probability.

\begin{figure}[t]
    \begin{centering}
    \includegraphics[width=6.5cm]{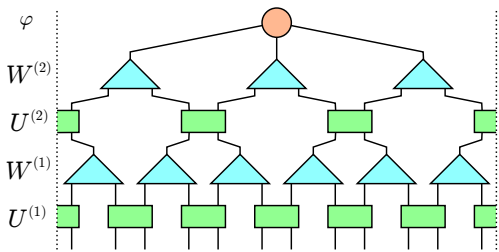}
    \caption{(Color online) Tensor network diagram for a binary MERA presenting the state $\ket{\Psi}$ of a translation invariant lattice $\mathcal{L}$ made of $L=12$ sites and with periodic boundary conditions. The tensors on each layer are identical and their labels are displayed to the left. The $U^{(n)}$ tensors (green rectangles) are unitary operators (acting top-to-bottom), $W^{(n)}$ (cyan triangles) are isometric, and $\varphi$ (red circle) is a normalized `wavefunction'. \label{fig_mera}}
    \end{centering}
\end{figure}

The main goals of this paper are to propose a variational Monte Carlo scheme for the MERA and to demonstrate its feasibility. We also discuss possible future applications. Let us briefly list some of the highlights of the approach. (i) In a lattice of size $L$, the sampled configurations correspond to an effective lattice of size $\mathcal{O}(\log(L))$; in this way, the cost of evaluating the expectation value of a local observable scales just as $\mathcal{O}(\log (L))$ and not as $\mathcal{O}(L)$ as in Refs.~\onlinecite{Schuch2008}, \onlinecite{Sandvik2007}. (ii) We employ the perfect sampling strategy of Ref.~\onlinecite{Ferris2011a}, thus avoiding the loses of efficiency in the Markov chain Monte Carlo of Refs.~\onlinecite{Schuch2008}, \onlinecite{Sandvik2007} due to equilibration and autocorrelation times. (iii) Variational parameters are optimized while explicitly preserving the unitary constraints that the tensors in the MERA are subjected to. This is accomplished by a steepest descent method within the set of unitary tensors, which is much more robust to statistical noise than the singular value decomposition methods employed in MERA algorithms without sampling.\cite{Evenbly2007}

We demonstrate the performance of our approach by computing an approximation to the ground state of a finite Ising chain with transverse magnetic field. For the binary MERA under consideration, sampling lowers the costs of elementary contractions from $\mathcal{O}(\chi^9)$ to $\mathcal{O}(\chi^5)$. We find that the resulting (approximate) ground state energy decreases as the number of samples is increased, thus obtaining a demonstration of principle of the approach. We also notice that the number of samples required to achieve a given accuracy increases as the transverse magnetic field approaches its critical value.

To our knowledge, this is the first instance of sampling-based optimization of a relatively complex tensor network. Previous similar optimizations included that of an MPS,\cite{Sandvik2007} which is a considerably simpler tensor network (with only three-legged tensors), and of tensor networks that under sampling break into smaller, simpler tensor networks (e.g. into MPS, single plaquette states, etc).\cite{Schuch2008, Changlani2009,Mezzacapo2009,Marti2010} In more complex tensor networks, such as MERA and PEPS, the optimization becomes much harder due to high sensitivity to statistical noise. Thus, for instance, Ref.~\onlinecite{Wang2011} spells out a full variational Monte Carlo approach for PEPS but uses an alternative method, not based on sampling, in order to optimize the tensors. Indeed, in Ref.~\onlinecite{Wang2011} sampling is only used to aid in the computation of expectation values. Here, instead, we use sampling both to optimize the MERA and to compute expectation values.

We emphasize, however, that our results only demonstrate a gain over optimization schemes based on exact contractions (i.e. without sampling) in the low accuracy regime, where only a relatively small number of samples are required. The specific MERA (namely binary MERA for a one-dimensional lattice) and low value of $\chi$ ($\chi=4$) considered here for illustrative purposes implies that the cost per sample is $\chi^{9}/\chi^{5} = \chi^4 = 256$ times smaller than an exact contraction. Recall that the statistical error decreases only as $\sqrt{N}$ with the number $N$ of samples. If more than $256$ samples are required in order to obtain a sufficiently accurate approximation of the exact contraction, then the sampling scheme may be overall less efficient than the exact contraction scheme.

The advantage of sampling over exact contraction schemes is expected to be more evident in MERA settings where the cost $\mathcal{O}(\chi^p)$ scales with a larger exponent $p$, and for larger values of $\chi$. In particular, we envisage that the method described in this paper, possibly with further improvements, will improve the range of applicability of MERA in two and higher dimensions.

The content of this paper is distributed in five more sections. In \Sec{sec_approach}, we discuss methods for sampling with the MERA. In \Sec{sec_optimize} we propose an optimization scheme using sampling techniques. In \Sec{sec_application} we benchmark the approach with the quantum Ising model. In \Sec{sec_discussion} we discuss future applications including extensions to higher dimensions and extracting long-range correlations, before concluding in \Sec{sec_conclusion}.

\section{Local expectation values with sampling}

\label{sec_approach}

In this section we explain how to use sampling in order to speed-up the computation of expectation values with the MERA. We present both complete and incomplete perfect sampling strategies, building on the proposals of Ref.~\onlinecite{Ferris2011a} for generic unitary tensor networks. We also discuss the importance of the choice of local basis in sampling. We start by reviewing some necessary background material on the MERA.

\subsection{MERA, expectation values and causal cones}

\begin{figure}[t]
  \begin{centering}
    \includegraphics[width=0.8\columnwidth]{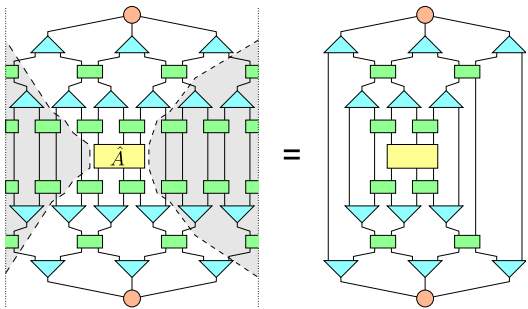}
    \caption{(Color online) Expectation value $\bra{\Psi} \hat{A} \ket{\Psi} = \bra{\Psi^{\mathcal{C}}} \hat{A} \ket{\Psi^{\mathcal{C}}}$ of the local operator $\hat{A}$ (yellow rectangle), which acts on (at most) three neighboring sites.  The flipped tensors, on the bottom half of the diagram, are the Hermitian conjugates of the respective tensors above. The `causal cone' is delimited by the dotted line in the left diagram, corresponding to $\bra{\Psi} \hat{A} \ket{\Psi}$. The tensors outside the causal cone cancel, significantly simplifying the diagram on the right, corresponding to $\bra{\Psi^{\mathcal{C}}} \hat{A} \ket{\Psi^{\mathcal{C}}}$. \label{fig_causal_cone}}
  \end{centering}
\end{figure}

The MERA\cite{Vidal2007b} is a variational wavefunction for ground states of quantum many-body systems on a lattice. The state $\ket{\Psi}$ of a lattice $\mathcal{L}$ made of $L$ sites is represented by means of a tensor network made of two types of tensors, called \emph{disentanglers} and \emph{isometries}. The tensor network is based on a real-space renormalization group transformation, known as \emph{entanglement renormalization}: disentanglers are used to remove short-range entanglement from the system, whereas isometries are used to coarse-grain blocks of site into single, effective sites. An example of a MERA on a periodic 1D lattice with $L = 12$ sites is depicted in \Fig{fig_mera}. This structure is called `binary' MERA because of the 2-to-1 course-graining transformation in each repeating layer. Ascending upwards in the figure, the disentanglers $U^{(n)}$ remove short-range entanglement in between each course-graining transformation, implemented by isometries $W^{(n)}$ until the remaining Hilbert space is small enough to deal with directly with some wavefunction $\varphi$.

\begin{figure}[t]
  \begin{centering}
    \includegraphics[width=\columnwidth]{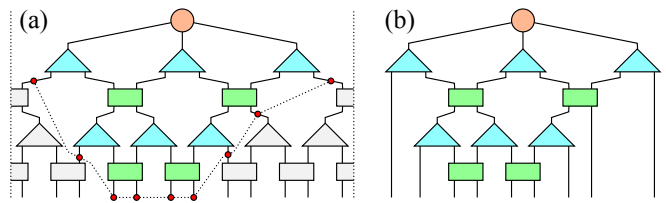}
    \caption{(Color online) States $\ket{\Psi}$ and $\ket{\Psi^{\mathcal{C}}}$ represented by the MERA. (a) Tensor network for the state $\ket{\Psi}$ of the original lattice $\mathcal{L}$. The causal cone is delimited by a discontinuous line. (b) Tensor network for the state $\ket{\Psi ^{\mathcal{C}}}$ of the effective lattice $\mathcal{L}^{\mathcal{C}}$. Sites further from the center effectively represent increasingly large length scales in the original lattice. \label{fig_sampled_wavefunction}}
  \end{centering}
\end{figure}

The MERA can also be viewed in the reverse --- starting from the top of \Fig{fig_mera}, we descend downwards in a unitary quantum circuit, adding (initially unentangled) sites in each layer. For instance, let us flow downwards in \Fig{fig_mera}. To a three-site system in state $\varphi$, we first add three additional unentangled sites, turning it into a six-site system; and later we add another six unentangled sites, producing the final twelve-site system. This unitary structure can be exploited when calculating the expectation value $\langle \hat{A} \rangle \equiv \bra{\Psi} \hat{A} \ket{\Psi}$ of a local operator $\hat{A}$ acting on a few neighboring sites. Specifically, all tensors not `causally' connected to the few sites supporting $\hat{A}$ cancel, as depicted in \Fig{fig_causal_cone}. The resulting diagram is significantly simpler and can be interpreted as the expectation value $\bra{\Psi^{\mathcal{C}}} \hat{A} \ket{\Psi^{\mathcal{C}}}$ of $\hat{A}$ for a state $\ket{\Psi^{\mathcal{C}}}$ of an effective lattice $\mathcal{L}^{\mathcal{C}}$ made of $O(\log(L))$ sites (see \Fig{fig_sampled_wavefunction}). We emphasize that, by construction,
\begin{equation}
\bra{\Psi} \hat{A} \ket{\Psi} = \bra{\Psi^{\mathcal{C}}} \hat{A} \ket{\Psi^{\mathcal{C}}}.
\end{equation}
Therefore, we can evaluate the expectation value $\bra{\Psi} \hat{A} \ket{\Psi}$ by contracting the tensor network corresponding to $\bra{\Psi^{\mathcal{C}}} \hat{A} \ket{\Psi^{\mathcal{C}}}$.
The numerical cost of performing this contraction grows linearly with the number of sites in $\mathcal{L}^{\mathcal{C}}$, and thus only logarithmically with the number of site $L$ in the original lattice $\mathcal{L}$.

The dimension of the Hilbert space after each course-graining transformation is an adjustable parameter, the \emph{bond dimension} $\chi$, which plays a central role in the present discussion. Increasing the bond dimension $\chi$ implies including a larger fraction of the original Hilbert space and leads to greater accuracy, but also requires greater computational resources. Optimization algorithms to approximate ground states, and to evaluate local expectation values and correlators, are present in the literature.\cite{Evenbly2007} The numerical cost of finding an expectation value or performing a single optimization iteration using the binary MERA scales as $\mathcal{O}(\chi^9 \log L)$ for a translation invariant system. For more complex MERA structures, such as those representing 2D lattices, the power of $\chi$ for the cost increases dramatically. For instance, the 2D MERA presented in Ref. \onlinecite{Evenbly2009} has a numerical cost of $\mathcal{O}(\chi^{16} \log L)$, which on current computers restricts $\chi < 8$. For many systems, this does not allow for enough entanglement to accurately describe the ground state, limiting the accuracy of the approach.

Here we hope to alleviate this problem by reducing the numerical cost as a function of $\chi$ using Monte Carlo techniques. We will find that the cost of a single sample scales as $\mathcal{O}(\chi^5)$ for binary 1D MERA, compared to the $\mathcal{O}(\chi^9)$ cost of the `exact' contraction.

\subsection{Monte Carlo sampling with the MERA}

\label{sec_sampling}

\begin{figure}[t]
  \centering
       \includegraphics[width=\columnwidth]{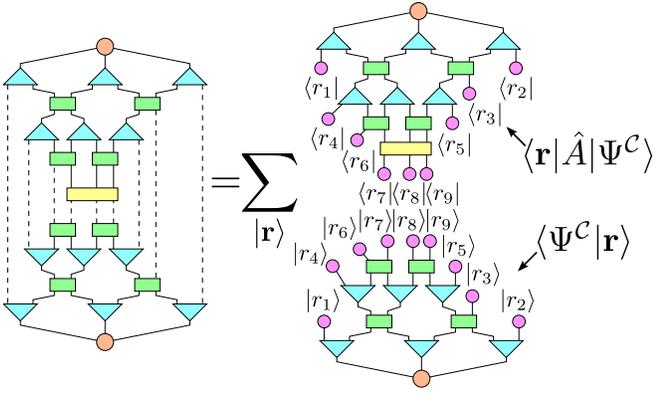}
        \caption{(Color online) Tensor network for $\bra{\Psi^{\mathcal{C}}}\hat{A}\ket{\Psi^{\mathcal{C}}}$. The dashed lines indicate the indices that will be sampled. On the right-hand-side we explicitly write the tensor contractions outside the causal cone as a sum over a complete, orthonormal set of `wavefunctions' $| \mathbf{r} \rangle = \ket{r_1} \otimes \ket{r_2} \otimes \dots$ (pink circles). Monte Carlo sampling will be performed over this set. Each term of the sum can be expressed as the product $\langle \Psi^{\mathcal{C}} | \mathbf{r} \rangle \langle \mathbf{r} | \hat{A} | \Psi^{\mathcal{C}} \rangle$. \label{fig_mera_structure}}
\end{figure}

Our goal is to compute the expectation value $\langle \hat{A} \rangle \equiv \bra{\Psi} \hat{A} \ket{\Psi}$ by contracting the tensor network corresponding to $\bra{\Psi^{\mathcal{C}}} \hat{A} \ket{\Psi^{\mathcal{C}}}$, see \Fig{fig_causal_cone}. The first step is to re-express the tensor network contraction as a summation over indices corresponding to the sites of the effective lattice, as shown in \Fig{fig_mera_structure}. We then get
\begin{equation}
  \langle \hat{A}\rangle = \sum_{\mathbf{r} \in \mathcal{R}} \bra{ \Psi^{\mathcal{C}}} \mathbf{r} \rangle \langle \mathbf{r} | \hat{A} \ket{ \Psi^{\mathcal{C}}}, \label{energy}
\end{equation}
where $\mathcal{R}$ is an orthonormal basis of product states $|\mathbf{r}\rangle = |r_1\rangle \otimes |r_2\rangle \otimes \dots$ on the effective lattice $\mathcal{L}^{\mathcal{C}}$.

We will approximately evaluate the sum in \Eqn{energy} by using Monte Carlo sampling over the states $|\mathbf{r}\rangle \in \mathcal{R}$. Notice that in the effective lattice, the sites away from the support of $\hat{A}$ have undergone one or more course-graining transformations. In other words, sites further from the center represent increasingly larger length-scales. Thus, sampling over sites of the effective lattice corresponds to sampling the system \emph{at different length scales}. This property is reminiscent of global or cluster updates used in existing Monte Carlo methods to solve critical systems.

A na\"ive scheme for approximating the sum in \Eqn{energy} would be to choose $|\mathbf{r}\rangle$ at random from $\mathcal{R}$, and evaluate $\langle \Psi^{\mathcal{C}} | \mathbf{r} \rangle \langle \mathbf{r} | \hat{A} | \Psi^{\mathcal{C}} \rangle$ according to the tensor networks in \Fig{fig_mera_structure}. The cost of obtaining a single sample scales as $\mathcal{O}(\chi^5 \log L)$. However, the statistical variance of a sampling scheme can be substantially reduced by implementing importance sampling --- in this case choosing configurations $|\mathbf{r}\rangle$ that are more likely. More precisely, sampling is implemented according to the wavefunction \emph{weight}, $P(\mathbf{r}) \equiv | \langle\Psi^{\mathcal{C}} | \mathbf{r} \rangle|^2$, which can be calculated efficiently as indicated in \Fig{fig_weight}. We can express \Eqn{energy} in a form more convenient for importance sampling,
\begin{equation}
  \langle \hat{A}\rangle = \sum_{\mathbf{r} \in \mathcal{R}} P(\mathbf{r}) A^{\mathcal{C}}(\mathbf{r}), \label{energy2pre}
\end{equation}
where
\begin{equation}
	A^{\mathcal{C}}(\mathbf{r}) \equiv \frac{\langle \Psi^{\mathcal{C}} |\mathbf{r} \rangle \langle \mathbf{r} |\hat{A} |\Psi^{\mathcal{C}} \rangle}{\langle \Psi^{\mathcal{C}} | \mathbf{r} \rangle\langle \mathbf{r} | \Psi^{\mathcal{C}} \rangle} = \frac{\langle \mathbf{r} | \hat{A} | \Psi^{\mathcal{C}} \rangle}{\langle \mathbf{r} | \Psi^{\mathcal{C}} \rangle}.
	\label{estimator}
\end{equation}

Note that because the MERA is normalized by construction, and therefore the weights sum to one, $\sum_{\mathbf{r} \in \mathcal{R}} P(\mathbf{r})=1$. However, during sampling only some subset $\tilde{\mathcal{R}}$ of the configurations are considered. One needs to renormalize the weights accordingly, so that the expectation value $\langle \hat{A} \rangle$ is approximated as
\begin{equation}
  \langle \hat{A} \rangle \approx \sum_{\mathbf{r} \in \tilde{\mathcal{R}}}  P(\mathbf{r}) A^{\mathcal{C}}(\mathbf{r})/\sum_{\mathbf{r} \in \tilde{\mathcal{R}}} P(\mathbf{r}).   \label{energy2}
\end{equation}

\subsubsection{Complete perfect sampling}

\begin{figure*}[t]
  \begin{centering}
    \includegraphics[height=6.0cm]{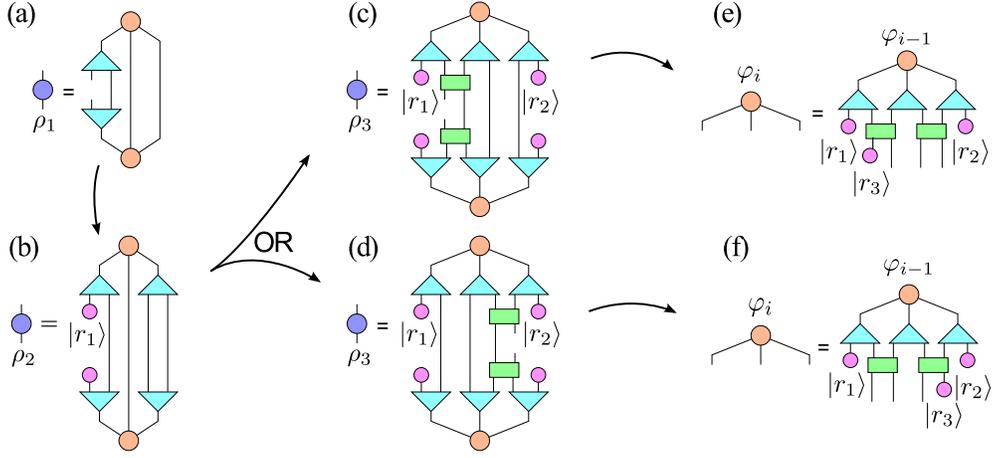}
    \caption{(Color online) Diagrams required to create a sample, sequentially sampling indices from the top to the bottom of the MERA. First the reduced density matrix, $\rho_i$ (blue circles), of an index is computed. We then randomly generate a state $|r_i\rangle$ according to the diagonal basis of $\rho_i$, and proceed to calculate $\rho_{i+1}$. In each `layer' of the MERA, we sample the outermost indices in (a,b) and \emph{one} of the inner indices in (c,d), depending on the physical location we are sampling. A projected, three-site wavefunction for the following layer is calculated in the appropriate diagram (e) or (f), from which the algorithm begins again with (a), until we have sampled all the indices indicated in \Fig{fig_mera_structure} or \ref{fig_alt_samp_scheme}. \label{fig_weight}}
  \end{centering}
\end{figure*}

In the case of MERA, and indeed any state that can be written as a unitary quantum circuit, a `perfect' sample can be generated according to the probability distribution $P(\mathbf{r})$ in a single sweep.\cite{Ferris2011a} This makes Markov Chain Monte Carlo unnecessary, simplifying the algorithm and eliminating a source of statistical error (i.e. autocorrelation effects). This is one advantage of this technique over other tensor network sampling methods in the literature.\cite{Sandvik2007, Schuch2008, Changlani2009, Mezzacapo2009, Wang2011, Marti2010}

The sample can be constructed by sampling just one index at a time. Beginning at the top layer of the MERA, and aiming to sample just the first index (left-most in \Fig{fig_mera_structure}~(b)), we can construct the one-site reduced density matrix $\rho_1$ by the tensor contraction in \Fig{fig_weight}~(b). The probability $P(r_1) = \langle r_1 | \rho_1 | r_1 \rangle$ can then be found \emph{for all possible} $|r_1\rangle$. A value of $|r_1\rangle$ is then randomly selected according to any complete basis of our choosing.

After this selection is made, we can then sample the next (top-most) index, according to the conditional weights $P(r_2 \, | \, r_1)$ as calculated by the diagram in \Fig{fig_weight}~(b) (we refer to Ref.~\onlinecite{Ferris2011a} for further details). We continue to sample the state of each site, until we have sampled every site.

Each of the diagrams in \Fig{fig_weight} can be calculated with cost $\mathcal{O}(\chi^5)$, while there are $\mathcal{O} (\log L) $ layers to the MERA. A single sample can therefore be generated with cost $\mathcal{O}(\chi^5 \log L)$, compared to $\mathcal{O}(\chi^9 \log L)$ for the exact contraction. So long as the number of samples $N$ is significantly less than $\chi^4$, Monte Carlo sampling will be faster than exact contraction.

In practice, we will perform $N \gg 1$ samples in order to get a good estimate of $\langle \hat{A} \rangle$. As the samples are completely uncorrelated, the variance of $A^{\mathcal{C}}(\mathbf{r})$
\begin{equation}
  \mathrm{Var} \bigl[ A^{\mathcal{C}} \bigr] \equiv \sum_{\mathbf{r}} P(\mathbf{r}) \left( A^{\mathcal{C}}(\mathbf{r})-\bar{A^{\mathcal{C}}}\right)^2
\end{equation}
can be used to estimate statistical error $\Delta A$,
\begin{equation}
  \Delta A \equiv \sqrt{\mathrm{Var}\bigl[ A^{\mathcal{C}} \bigr] / N}, \label{std_error}
\end{equation}
while $\mathrm{Var} \bigl[ A^{\mathcal{C}} \bigr]$ is itself upper bounded by the variance of the operator to be measured,
\begin{equation}
  \mathrm{Var} \bigl[ A^{\mathcal{C}} \bigr] \le \langle \hat{A}^2 \rangle - \langle \hat{A} \rangle^2. \label{std_error2}
\end{equation}
It is easy to show that the upper bound is saturated when sampling in the diagonal basis of $\hat{A}$. However, the actual value of this variance is dependent on the choice of basis $\{|\mathbf{r}\rangle\}$, as well as the state being sampled. Therefore, it is worth spending some effort in order to minimize this quantity.

\subsubsection{Incomplete perfect sampling}

\begin{figure}[t]
  \begin{centering}

    \includegraphics[width=0.95\columnwidth]{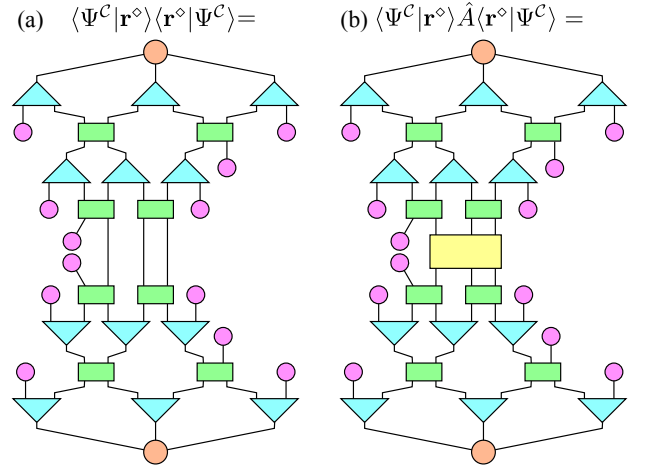}
    \caption{(Color online) Tensor network diagrams for the quantities (a)~$\langle \Psi^{\mathcal{C}} | \mathbf{r}^{\diamond} \rangle\langle \mathbf{r}^{\diamond} | \Psi^{\mathcal{C}} \rangle = P(\mathbf{r})$ and (b)~$\langle \Psi^{\mathcal{C}} |   \mathbf{r}^{\diamond} \rangle \hat{A} \langle \mathbf{r}^{\diamond} | \Psi^{\mathcal{C}} \rangle$ with incomplete sampling. \label{fig_alt_samp_scheme}}
  \end{centering}
\end{figure}

The statistical noise is caused by Monte Carlo sampling of the tensor contraction, and it stands to reason that sampling \emph{less} indices in the tensor network diagram (see \Fig{fig_mera_structure}) would reduce the statistical error. Indeed, it is possible to exactly contract the three physical indices at the lowest level (those connected to operator $\hat{A}$) while keeping the computational cost at $\mathcal{O}(\chi^5 \log L)$, as depicted in \Fig{fig_alt_samp_scheme}. That is, we sample over configurations $\mathbf{r}^{\diamond} \in \mathcal{R}^{\diamond}$ of the effective lattice $\mathcal{L}^{\mathcal{C}}$ \emph{minus} the three central sites on which $\hat{A}$ is supported.

This method effectively generates (unnormalized) three-site wave functions $\ket{ \varphi_0 (\mathbf{r}^{\diamond}) }\equiv \langle \mathbf{r}^{\diamond} | \Psi^{\mathcal{C}} \rangle$ from the reduced density operator (as seen in \Fig{fig_weight} (e,f)). Conversely, the reduced density matrix for the three central sites is  $\sum
_{\mathbf{r}^{\diamond}} | \varphi_0 (\mathbf{r}^{\diamond}) \rangle\langle \varphi_0 (\mathbf{r}^{\diamond}) |$. Importance sampling is now achieved by selecting $\mathbf{r}^{\diamond}$ according to the weight
\begin{equation}
	P(\mathbf{r}^{\diamond}) \equiv \langle \Psi^{\mathcal{C}} | \mathbf{r}^{\diamond} \rangle\langle \mathbf{r}^{\diamond} | \Psi^{\mathcal{C}} \rangle = \langle \varphi_0(\mathbf{r}^{\diamond}) | \varphi_0 (\mathbf{r}^{\diamond}) \rangle.
\end{equation}
In turn, perfect sampling of the sites proceeds exactly as before, but it stops before the three central sites, which are not sampled. We define the estimator
\begin{equation}
A^{\diamond}(\mathbf{r}^{\diamond}) \equiv \frac{\langle \Psi^{\mathcal{C}} |\mathbf{r}^{\diamond} \rangle \hat{A}\langle \mathbf{r}^{\diamond}| \Psi^{\mathcal{C}} \rangle}{\langle \Psi^{\mathcal{C}} | \mathbf{r}^{\diamond} \rangle\langle \mathbf{r}^{\diamond} | \Psi^{\mathcal{C}} \rangle} = \frac{\langle \varphi_0(\mathbf{r}^{\diamond}) | \hat{A} | \varphi_0 (\mathbf{r}^{\diamond}) \rangle}{\langle \varphi_0(\mathbf{r}^{\diamond}) | \varphi_0 (\mathbf{r}^{\diamond}) \rangle}, \label{EofS}
\end{equation}
whose expectation value obeys,
\begin{equation}
  \langle \hat{A}\rangle = \sum_{\mathbf{r}^{\diamond} \in \mathcal{R}^{\diamond}} P(\mathbf{r}^{\diamond}) A^{\diamond}(\mathbf{r}^{\diamond}),
\end{equation}
in analogy with \Eqn{energy2pre}. Notice that, by construction, the statistical variance $\mathrm{Var} \bigl[ A^{\diamond} \bigr]$ is smaller or equal than $\mathrm{Var} \bigl[ A^{\mathcal{C}} \bigr]$ in \Eqn{std_error2}, and therefore the numerical accuracy is increased without affecting the computational cost.

\subsubsection{Diagonal basis selection}

\label{sec_diagonal}

In general, we are free to choose any complete basis $\mathcal{R}$ (or $\mathcal{R}^{\diamond}$) from which to draw individual samples. A good choice is one that produces a small statistical error, Eqs. (\ref{std_error},\ref{std_error2}). Intuitively, the goal of importance sampling is to decrease the statistical variance by choosing configurations $\mathbf{r}$ (or $\mathbf{r}^{\diamond}$) with a large overlap with the state $\ket{\Psi^{\mathcal{C}}}$. With this in mind, one could aim to maximize the `average' weight,
\begin{equation}
  \overline{P} \equiv \sum_{\mathbf{r}} P(\mathbf{r})^2 / \sum_{\mathbf{r}} P(\mathbf{r}).
  \label{eq:weight}
\end{equation}
It is easy to show that for a given quantum probability distribution, specified by a density matrix, the above quantity is maximized in the diagonal basis of the density matrix. Inspired by this fact, here we choose to sample site $i$ in the basis in which the reduced density matrix $\rho_i$ is diagonal. The $\chi \times \chi$ density matrices $\rho_i$, calculated in \Fig{fig_weight}, can be diagonalized with cost $\mathcal{O}(\chi^3)$. Note that the chosen basis will depend on previously sampled sites, and that the resulting sampling basis is still a complete, orthonormal basis of product states.

We have found that this approach can radically increase the average value of the weight, \Eqn{eq:weight} with the effect becoming stronger for larger systems and values of $\chi$. More importantly, we find that the statistical variance in the observables is decreased (see \Sec{sec_application_expectation} and \Fig{fig_variance}).

This technique to select the sampling basis is not specific to unitary tensor networks nor perfect sampling methods, and could thus be of benefit to other variational quantum Monte Carlo algorithms.

\subsubsection{Operators that decompose as sum of local terms}

Finally, we may wish to compute the expectation value of an operator that is the sum of local terms, such as a Hamiltonian made of nearest-neighbor interactions:
\begin{equation}
  \hat{H} = \sum_i \hat{H}_i \label{local_H}.
\end{equation}
In this case we sample each local term $\hat{H}_i$ as indicated previously, noticing that the causal cone of each $\hat{H}_i$ depends on the location of the sites of lattice $\mathcal{L}$ where the local operator is supported. One can either choose to (uniformly) sample the position $i$ in the lattice, or systematically sweep through all the positions $i$. A complete sweep, where each site $i$ is visited once, costs $\mathcal{O}(\chi^5 L\log L)$.

\section{Energy minimization with sampling}

\label{sec_optimize}

\begin{figure*}[t]
  \begin{centering}
    \includegraphics[width=1.8\columnwidth]{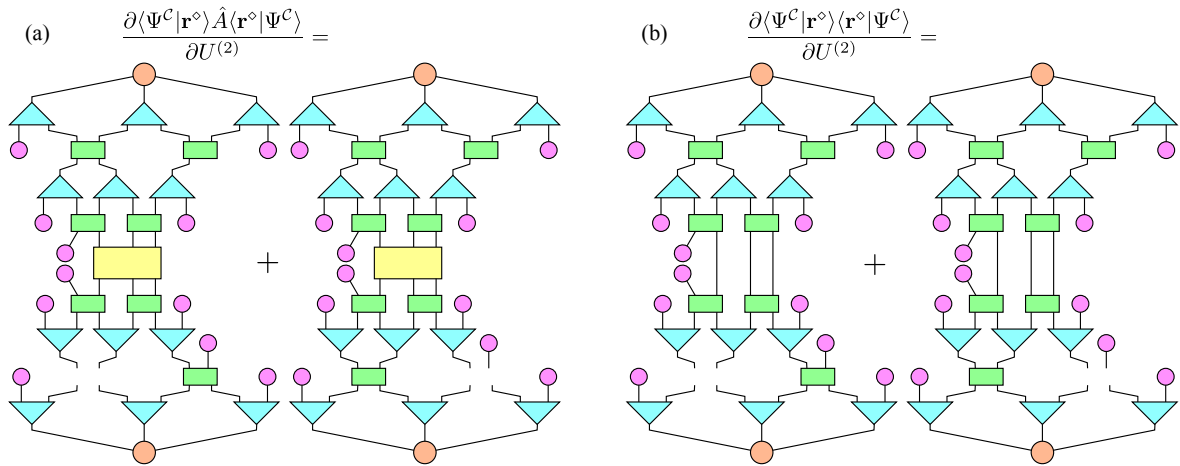}
    \caption{(Color online) Examples of tensor network diagrams to find the derivatives of $\langle \Psi^{\mathcal{C}} |\mathbf{r}^{\diamond} \rangle \hat{A} \langle \mathbf{r}^{\diamond} | \Psi^{\mathcal{C}} \rangle$ and $\langle \Psi^{\mathcal{C}} | \mathbf{r}^{\diamond} \rangle\langle \mathbf{r}^{\diamond} | \Psi^{\mathcal{C}} \rangle$  with respect to the tensor $U^{(2)\dag}$ (i.e. the transpose of those in \Eqn{derivative}). The derivative with respect to a particular tensor is given by it's \emph{environment}, or the contraction of all the other tensors in the diagram. The chain rule produces multiple terms where a tensor appears more than once. \label{fig_derivatives}}
  \end{centering}
\end{figure*}

In order to find an approximation to the ground state, we need to minimize the energy of the MERA. The direction of steepest \emph{ascent} is given by the complex derivative with respect to the conjugate~\cite{Hjorungnes2007} of each element of each tensor. Inserting \Eqn{estimator} (or \Eqn{EofS}) into \Eqn{energy2} and differentiating gives
\begin{eqnarray}
	&&\frac{\partial \langle \hat{H} \rangle}{\partial U^{(n)\ast}} = \sum_{i} \frac{\partial \langle \hat{H}_i \rangle}{\partial U^{(n)\ast}}, \\
	    &&\frac{\partial \langle \hat{H}_i \rangle}{\partial U^{(n)\ast}} =
    \sum_{\mathbf{r}} P(\mathbf{r})  \biggl[ \frac{1}{P(\mathbf{r})} \frac{\partial \langle \Psi^{\mathcal{C}}|  \mathbf{r} \rangle \langle \mathbf{r} |\hat{H}_i |\Psi^{\mathcal{C}} \rangle}{\partial U^{(n)\ast}}  \nonumber \\
    && ~~~~~~~~~~~~~~~~~~~~~~~~~~~~~~~~~ - \frac{\langle \hat{H}_i \rangle}{P(\mathbf{r})} \frac{\partial \langle \Psi^{\mathcal{C}} | \mathbf{r} \rangle \langle \mathbf{r} | \Psi^{\mathcal{C}} \rangle}{\partial U^{(n)\ast}} \biggr] , \label{derivative}
\end{eqnarray}
where the derivative with respect to a tensor is element-wise, and similar expressions hold for $W^{(n)}$ and $\varphi$. The derivatives on the right-hand-side of \Eqn{derivative} can be found by using the usual rules for calculating derivatives in the diagrams, see Figs. \ref{fig_alt_samp_scheme} and \ref{fig_derivatives}.\footnote{Notice that the second term in the brackets of \Eqn{derivative} arises from the change in the normalization of the wave function. Although we are dealing with unitary/isometric tensors which ensure $\langle \Psi^{\mathcal{C}} | \Psi^{\mathcal{C}} \rangle = 1$, small changes in arbitrary directions may break the unitarity and modify the norm. Interestingly, even when projecting into the unitary tangent space (see below) where this term averages to zero, its inclusion is important to reduce the sampling error --- sometimes by several orders of magnitude.} In practice $\langle \hat{H}_i \rangle$ and $\frac{\partial \langle \hat{H}_i \rangle}{\partial U^{(n)\ast}}$ will be estimated simultaneously by sampling.

Multiple approaches are possible for updating the MERA to minimize the energy and find a good approximation to the ground state. One approach often used iteratively optimizes each tensor in the MERA according to the following algorithm\cite{Evenbly2007}.
\begin{eqnarray}
  u s v & = & - \frac{\partial \langle \hat{H} \rangle}{\partial U^{(n)\ast}} \nonumber \\
  U^{(n)} & \rightarrow & u v
\end{eqnarray}
In the above $u$ and $v$ are unitary, while $s$ is diagonal and positive, thus representing the singular value decomposition (SVD) of the derivative. This algorithm finds the unitary tensor $U^{(n)}$ that minimizes the trace value of its product with the above, called environment--- with the requirement that $\hat{H}$ is negative-semidefinite. Similar steps apply to $W^{(n)}$ and $\varphi$, where the SVD ensures that they remain isometric or normalized, respectively.

Unfortunately, the above scheme is extremely sensitive to the statistical noise inherent to Monte Carlo sampling, and results in very poor optimization. Ideally, we would prefer a method in which the statistical noise is able to average out over many iterations.

The most obvious scheme satisfying this requirement is straightforward steepest descent. Again, one must ensure that the tensors obey the unitary/isometric constraints characteristic of the MERA, so one can utilize the SVD to find the unitary tensor closest (with respect to the $L_2$--norm) to the usual downhill update. With this method, the $i$th step is given by
\begin{eqnarray}
  u s v & = & U^{(n)} - \mu_t \frac{\partial \langle \hat{H} \rangle}{\partial U^{(n)\ast}}, \nonumber \\
  U^{(n)} & \rightarrow & u v,
\end{eqnarray}
where $\mu_t$ is a number modulating the size of change at the step $t$.

In this paper we avoid using the SVD entirely by explicitly remaining in the unitary subspace, along the lines of Ref.~\onlinecite{Abrudan2008}. We define the tangent vector $G_U^{(n)}$ as the derivative projected onto the tangent space of all unitaries located about $U^{(n)}$,
\begin{equation}
  G_U^{(n)} = \frac{\partial \langle \hat{H} \rangle}{\partial U^{(n)\ast}} - U^{(n)} \frac{\partial \langle \hat{H} \rangle}{\partial U^{(n)\ast}}^{\dag} U^{(n)}.
\end{equation}
The matrix $U^{(n)} - \mu_t G^{(n)}_U$ is within $\mathcal{O}(\mu_t^2)$ of a unitary matrix. Noting that $U^{(n) \dag} G^{(n)}_U$ is anti-Hermitian, then the update
\begin{equation}
U^{(n)} \rightarrow U^{(n)} \exp\left[-\mu_t \left( U^{(n) \dag} \frac{\partial \langle \hat{H} \rangle}{\partial U^{(n)\ast}} - \frac{\partial \langle \hat{H} \rangle}{\partial U^{(n)\ast}}^{\dag} U^{(n)}\right)\right]
\end{equation}
both travels in the direction of the tangent vector while $U^{(n)}$ remains precisely unitary. The same approach works for $n \times m$ isometric matrices, taking care that $n \le m$, with computational cost scaling similarly to the SVD approach as $\mathcal{O}(n^2 m)$ (see Appendix).

The performance of the algorithm is highly dependent on the behaviour of $\mu_t$, as well as the number of Monte Carlo samples, $N_t$, taken in each step. Simple schemes will keep $\mu_t$ and $N_t$ constant, which is the approach we take here. On the other hand, one may choose to increase $N_t$ with $t$ so that harmful noise is reduced when approaching the optimal solution; or to decrease $\mu_t$ with $t$ for much the same reason; or a combination of both.

\section{Benchmark: critical quantum Ising chain}

\label{sec_application}

In this section we demonstrate the above techniques with the well-known transverse-field quantum Ising model,
\begin{equation}
  \hat{H} = - \sum_i \hat{\sigma}^z_i \hat{\sigma}^z_{i+1} + h \hat{\sigma}^x_i. \label{ising}
\end{equation}
Such a Hamiltonian can be expressed as a sum of nearest-neighbour terms $\hat{H}_i$, such that
\begin{equation}
  \hat{H}_i = - \hat{\sigma}^z_i \hat{\sigma}^z_{i+1} - h/2 \bigl(\hat{\sigma}^x_i + \hat{\sigma}^x_{i+1}\bigr). \label{ising2}
\end{equation}
We will pay particular attention to the region around the critical point at $h = 1$, which is the most demanding computationally. For concreteness, we use a three-layer binary MERA with periodic boundary conditions, resulting in a lattice of 24 sites in the bottom of the MERA structure. However, each of these sites corresponds to a block of $3$ physical spins, making a total of 72 spins. We choose this blocking so that for $\chi \le 2^3 = 8$ the bond dimension only ever decreases when ascending through the MERA. In what follows, we employ incomplete perfect sampling where three sites at the bottom of the MERA are contracted exactly.

\subsection{Expectation values}

\label{sec_application_expectation}

\begin{figure}
  \begin{centering}
    \includegraphics[width=0.85\columnwidth]{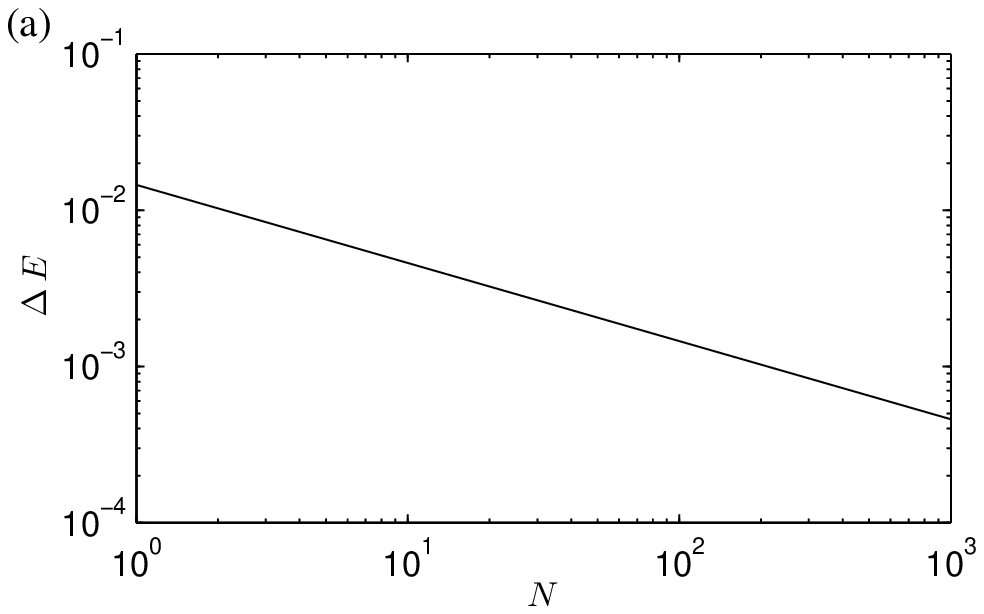}
    \includegraphics[width=0.85\columnwidth]{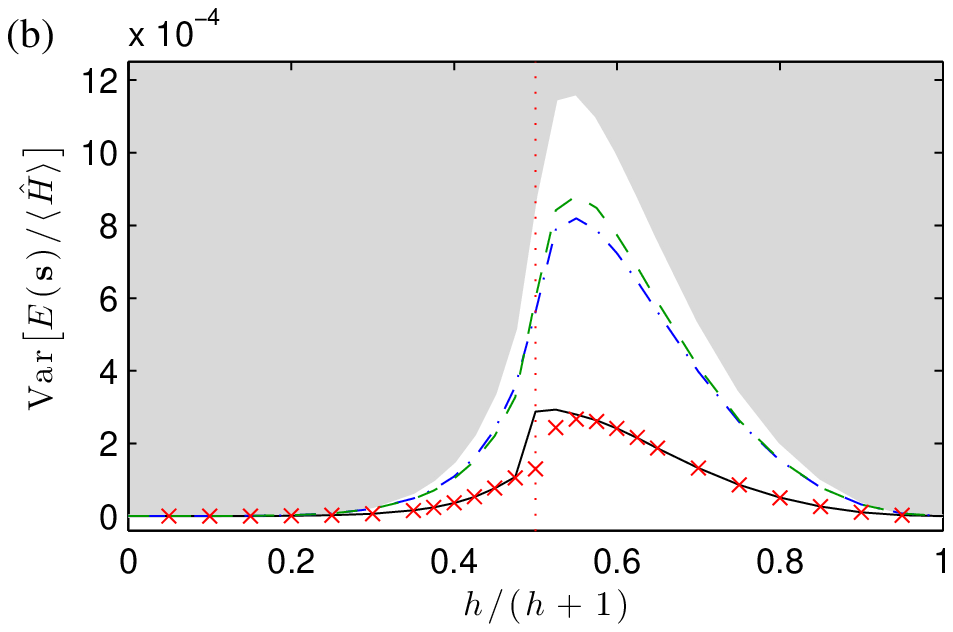}
    \includegraphics[width=0.85\columnwidth]{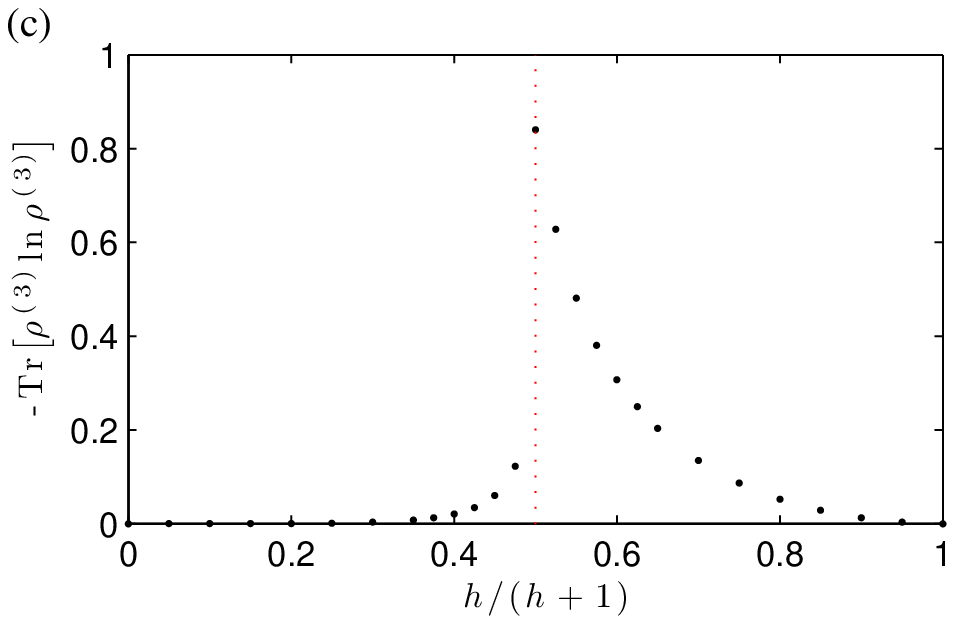}
  \end{centering}
  \caption{(Color online) (a) The statistical error of the energy per site at $h=1$ as a function of the number of samples $N$ follows the classic $N^{-1/2}$ scaling. (b) Variance of the energy estimator (normalized to the expectation value $\langle\hat{H}\rangle$) for optimized MERAs with bond-dimension $\chi = 8$ for various values of $h$ (red crosses). Estimates of the statistical uncertainty are smaller than the symbols. The grey area is eliminated by inequality \Eqn{std_error2}, bounded by the variance of $\hat{H}$. For comparison, we include the variances expected from several hypothetical samplings of $\rho^{(3)}$. Variances from sampling the spins in the $z$ (or $x$) basis is indicated by a blue, dash-dot (or green, dashed) line. Our numerical results (red crosses) show remarkable similarity with sampling in the diagonal basis of $\rho^{(3)}$ (black, solid line). (c)~The entanglement entropy of $\rho^{(3)}$. In (b) and (c), the critical point at $h = 1$ is indicated with a red dotted vertical line. } \label{fig_variance}
\end{figure}

We now analyze the effectiveness of extracting expectation values from the MERA using Monte Carlo sampling. For perfect sampling techniques, the accuracy can be easily extracted from the variance using \Eqn{std_error}. The scaling of the error in the energy, $\Delta E \propto N^{-1/2}$, is shown explicitly for the critical ($h=1$) system in \Fig{fig_variance}~(a). The variance of the energy estimator $E$ as a function of $h$ is shown in \Fig{fig_variance}~(b). Here we have used MERA wavefunctions previously optimized using standard techniques\cite{Evenbly2007} \emph{without} Monte Carlo sampling, that is, sampling is only employed to extract the expectation values.

Notice that the variance is maximal near the critical point at $h = 1$. As the Monte Carlo code effectively samples wavefunctions from the reduced three-site (i.e. nine-spin) density operator, one would expect the energy variance to increase with the amount of entanglement in the system. For reference we have included the entropy of the three-site density matrix $\rho^{(3)}$ in \Fig{fig_variance}~(c). This entropy mostly\footnote{There may be very small contributions to the entropy resulting from the anisotropy of the MERA. Also note that $\hat{H}$ and $\rho^{(3)}$ share a $\mathbb{Z}_2$ symmetry which is broken by the MERA wavefunction for $h < 1$, and the \emph{exact} ground state should have $S = \ln 2$ entanglement entropy as $h \rightarrow 0$.} corresponds to the entanglement entropy of three sites with the remainder of the system. We see a strong correlation between the amount of entanglement and the size of the variance of the energy estimator $E$.

Let us emphasize that \Fig{fig_variance}~(b) shows that our scheme performs significantly better than directly sampling $\rho^{(3)}$ in either the $x$ or $z$ spin basis. The measured variances are very similar to a diagonal sampling of $\rho^{(3)}$ (i.e. in its diagonal basis). This indicates that the sampling scheme is performing as intended in \Sec{sec_diagonal}.

In general, an accurate representation of wavefunctions with greater amounts of entanglement will require greater bond dimension $\chi$. These results suggest that the statistical variance generated by this scheme will also increase with the entanglement. To achieve a certain precision in the expectation value of local observables, the number of required samples grows with this variance, and thus with the amount of entanglement and with the minimum suitable value of $\chi$. Therefore, although a \emph{single} sample has cost $\mathcal{O}(\chi^5)$, the \emph{total} cost to obtain a certain precision may have some additional dependence on $\chi$. Nevertheless, no additional dependence was clearly manifest in our simulations at fixed $h$.


\subsection{Optimization}

\begin{figure}
  \begin{centering}
    \includegraphics[width=0.85\columnwidth]{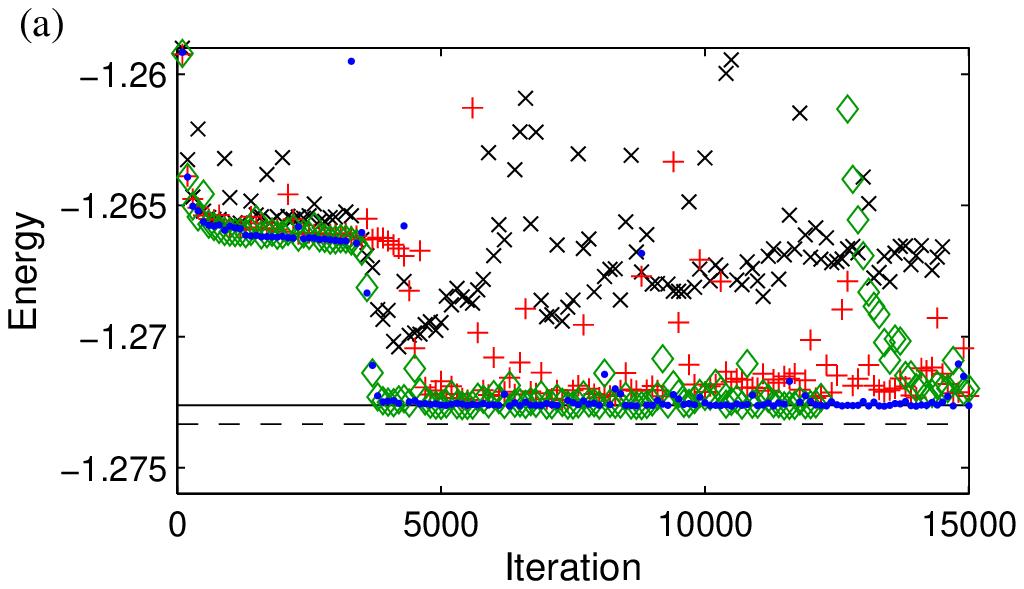}
    \includegraphics[width=0.85\columnwidth]{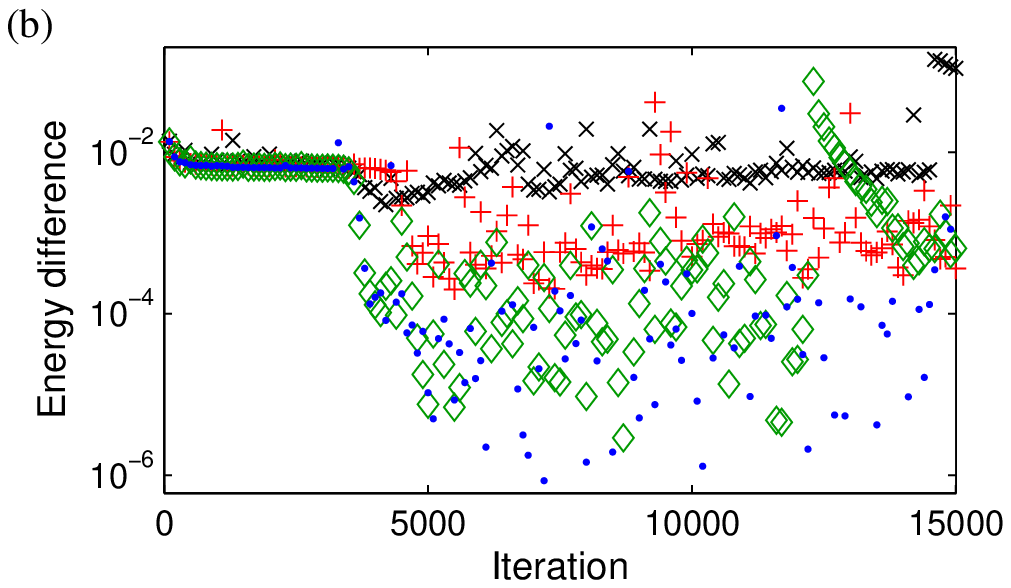}
  \end{centering}
  \caption{(Color online) (a) Energy of a wavefunction with $\chi = 4$ during optimization. Each iteration is update using $N$ sweeps, where $N$ is $1,2,4,8$ in the black crosses, red pluses, green diamonds and blue points respectively. Every 100 iterations, we calculate the exact energy corresponding to the current wavefunction, which is plotted here. The solid horizontal line indicates the energy of an optimized $\chi=4$ MERA using exact contractions and steepest descent (which remains $\approx 7 \times 10^{-4}$ above the the true ground state energy, indicated by the dashed line). The simulation converges for large $N$, but $\chi$ may need to increase for greater accuracy. (b) Difference to the above solid line plotted on a logarithmic scale. The difference reduces with increasing $N$, and although statistical fluctuations are decreasing, they remain evident on the logarithmic scale.} \label{fig_optimize}
\end{figure}

Finally, we combine Monte Carlo sampling with our unitary-subspace steepest descent algorithm to obtain optimized wavefunctions.  In \Fig{fig_optimize} we plot the energy of the MERA during the optimization process at $h = 1$ and $\chi = 4$, where the simulation progresses through a range of different number of sweeps $N$ per optimization step. In all cases the step size is fixed at $\mu_t = 0.1$. We observe that increasing $N$ improves the quality of the optimized wavefunction and for large values of $N$ the simulation tends to converge towards the same energy obtained with exact contractions, as expected.

Like all tensor network optimizations, care must be taken to ensure the wavefunction has fully converged to the lowest energy state. For instance, in \Fig{fig_optimize} (a) we see a plateau in energy before around the 4000th iteration that could be mistaken for convergence (whereas the simulation is actually navigating a stiff region, i.e. a long narrow valley in the energy landscape). Non-deterministic features due to statistical fluctuations can also be seen --- such as the sudden increase of energy of the $N =4$ simulation around the 13000th iteration.

Beyond this, accuracy could be improved by increasing $\chi$. It should be noted that we have observed that the steepest descent method (with either exact contractions or sampling) will not always produce wavefunctions of the same quality as the SVD method as it may be more susceptible to local minima or extreme stiffness. However, accuracy can still be systematically improved by increasing $\chi$.

\begin{figure}
  \begin{centering}
    \includegraphics[width=0.8\columnwidth]{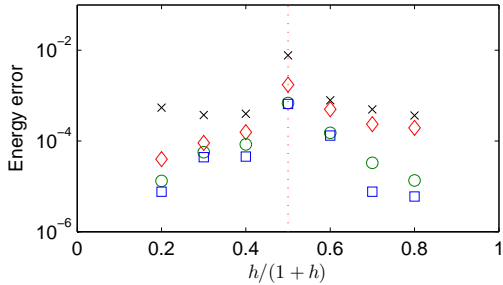}
  \end{centering}
  \caption{(Color online) The difference between the final optimized energy and the true ground state for a variety of field strengths $h$. The markers indicate the number of Monte Carlo sweeps taken between updates, $N = 1,4,16,64$ for the black cross, red diamond, green circle and blue square, respectively. There is a clear trend for improved ground state energy as $N$ increases, and away from the critical point at $h = 1$ (vertical red dotted line).} \label{fig_optimize_vs_h}
\end{figure}

In the previous section we noted that the statistical uncertainty peaked around the critical point at $h = 1$, where the entanglement is maximal, and one might expect the optimizations to be most difficult around this point. Plotted in \Fig{fig_optimize_vs_h} is the difference in energy between our wavefunctions and the exact, analytic solution for a range of $h$. We observe that the error decreases away from the critical point, and that there is a clear relationship between the quality of the wavefunction and the number of samples per step, $N$.

There are several possible limiting factors in variational Monte Carlo optimizations of MERA wavefunctions. One must balance the cost of increasing $N$, $\chi$ and the total number of iterations, to produce results of the desired accuracy. On top of this, the ansatz presents a complicated optimization landscape and one must be careful not to be stuck in local minima.

There is much scope to improve on the above optimization scheme by using more sophisticated approaches. Most obviously, the step-size $\mu_t$ and number of samples $N_t$ performed in each iteration could be adjusted as the simulation progresses. For example by choosing the step size to decrease as $\mu_t \propto 1 / t^{\alpha}$, with $N$ fixed and $0 < \alpha \le 1$ we are guaranteed convergence to some local minimum\cite{Harju1997}. Equivalently, the noise could be reduced at each step by increasing $N \propto t^{\beta}$, or some combination of both.

In Ref.~\onlinecite{Sandvik2007} it was found that using just the sign of the derivative, as well as properties resulting from translational invariance, was sufficient for optimizing a periodic MPS with Monte Carlo sampling. Other approaches existing in the literature may result is significant gains, though it should be noted that approaches requiring the second derivative or Hessian matrix would increase the order of the numerical cost as a function of $\chi$, and would have to be made robust to statistical noise.

\section{Future applications}

\label{sec_discussion}

There are several situations where it is most natural to use Monte Carlo sampling to speed-up MERA algorithms. Let us briefly review them.

\subsection{Different MERA structures}

\label{sec_other_structures}

In this work we have considered in detail the binary MERA for a 1D lattice with translation invariance. However, even in a translationally invariant, 1D lattice, one has freedom to choose between a large variety of entanglement renormalization schemes, leading to MERA structures with different configurations of isometries and disentanglers.

\begin{figure}[t]
  \begin{centering}
    \includegraphics[width=5cm]{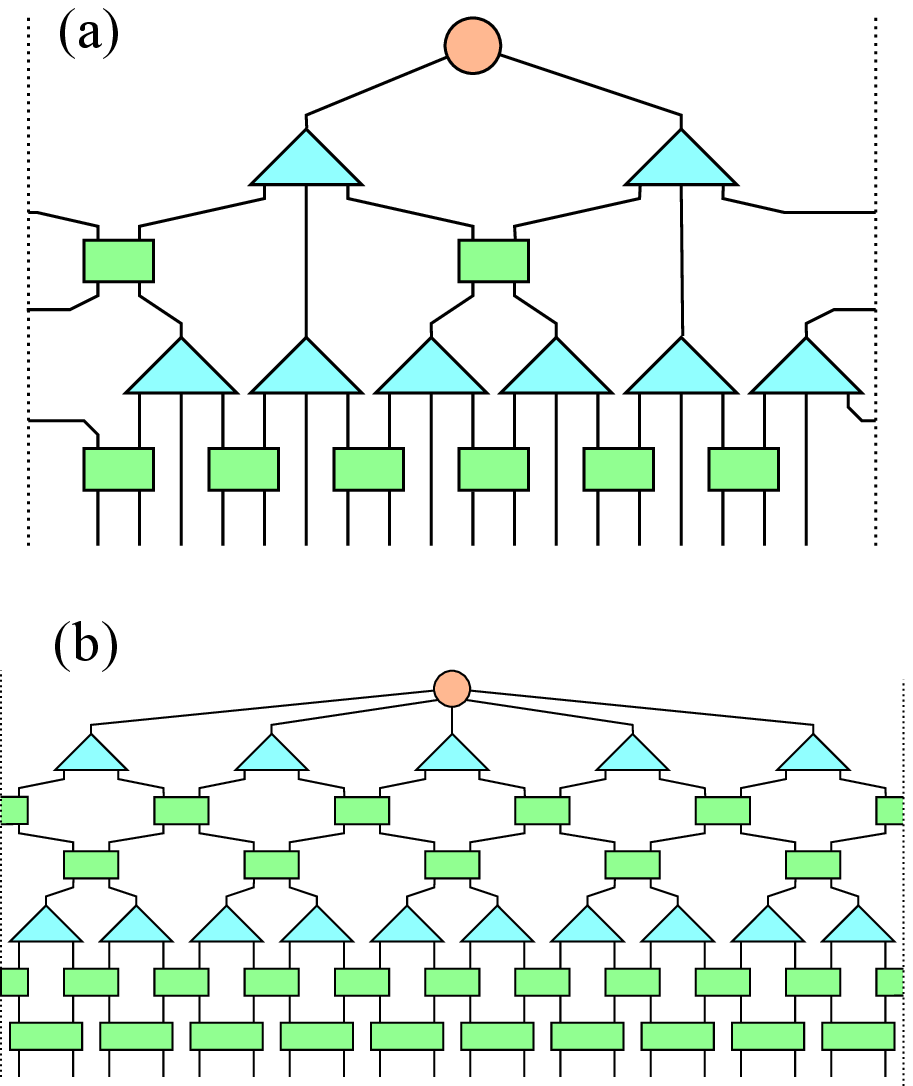}\includegraphics[width=3.5cm]{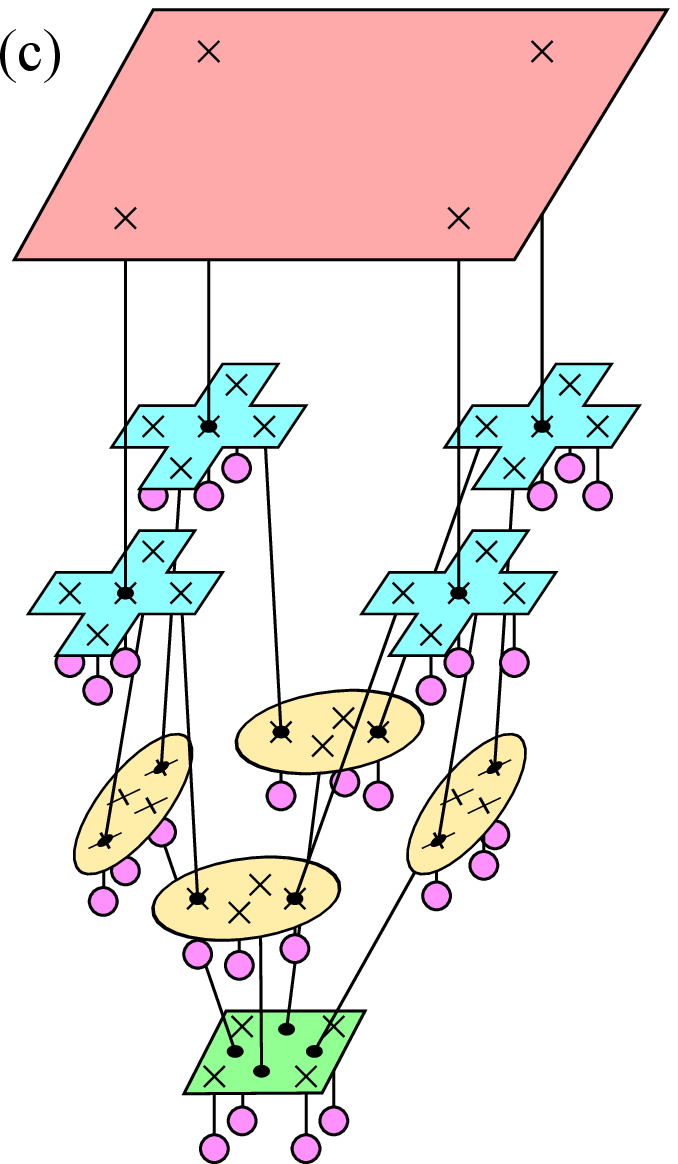}
    \caption{(Color online) (a) The ternary MERA structure has isometries transforming 3 sites into 1. (b) This two-layer binary MERA structure has added disentanglers which may help represent long range entanglement. (c) Sampling scheme for the lowest layer of the $3\!\!\times\!\!3$-to-$1$ MERA described in Ref.~\onlinecite{Evenbly2009}. \label{fig_other_meras}}
  \end{centering}
\end{figure}

The ternary MERA, shown in \Fig{fig_other_meras}~(a), has a narrower causal cone, with a width of just two sites, and the traditional algorithms for optimizing it have a cost that scales as $\mathcal{O}(\chi^8)$. As a result, the ternary MERA is sometimes favored over the binary MERA. Note that this does not necessarily translate into an improved accuracy in expectation values --- the ternary MERA is in general less accurate than binary MERA for the same value of $\chi$.

It is interesting to note that, in the ternary MERA, the perfect sampling algorithm presented here again has cost $\mathcal{O}(\chi^5)$, while Markov chain Monte Carlo, as well as expectation value and environment estimation, is possible with cost $\mathcal{O}(\chi^4)$. It is unclear whether after including autocorrelation effects the Markov chain method performs better, similar, or worse overall compared to the perfect sampling algorithm.

Another possible 1D MERA includes two layers of disentanglers to account for entanglement over larger distances, as depicted in \Fig{fig_other_meras}~(b). This MERA has a causal cone that is five sites wide, and traditional algorithms would have numerical cost $\mathcal{O}(\chi^{12})$ and require memory $\mathcal{O}(\chi^{10})$ --- limiting $\chi$ to rather small values. However, a sampling technique will only require $\mathcal{O}(\chi^7)$ time per sample and $\mathcal{O}(\chi^5)$ memory overall --- a huge  saving. Note that the power roughly halves when we change from an exact contraction which effectively rescales density matrices from higher layers to lower, to a Monte Carlo scheme which samples wavefunctions from this distribution.

The scaling of computational cost in $\chi$ in 2D lattices is even more challenging, mostly because the width of the causal cone (or number of indices included in a horizontal section of the causal cone) is much larger. Once again, sampling wavefunctions will require roughly square-root the number of operations (and memory) needed to calculate the exact reduced-density matrix. For instance, in the $3\!\!\times\!\!3$-to-$1$ MERA presented in Ref.~\onlinecite{Evenbly2009}, the cost of an exact contraction scales as $\mathcal{O}(\chi^{16} \log L)$, while with Monte Carlo sampling it is possible with just $\mathcal{O}(\chi^8 \log L)$ operations per sample (depicted in \Fig{fig_other_meras}~(c)). Memory might be a limiting factor in 2D MERA algorithms, while the temporary memory required for this algorithm is less than that to store the disentanglers and isometries.

\subsection{Long-range correlations}

Another challenge with MERA calculations is the numerical cost of long-range correlations. Take for instance the two-site operator $\langle \hat{A}_i \hat{B}_j\rangle$, for arbitrary sites $i$ and $j$. The cost required to contract the corresponding tensor network within the binary MERA scheme can scale as much as $\mathcal{O}(\chi^{12})$ --- significantly more than the $\mathcal{O}(\chi^9)$ cost for neighbor and next-nearest-neighbor correlations.

\begin{figure}[t]
  \begin{centering}
   \includegraphics[width=\columnwidth]{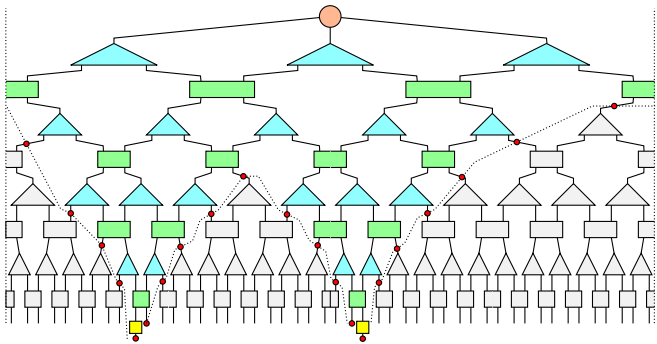}
    \caption{(Color online) Causal cone of two operators separated by 17 sites in a 4 layer binary MERA. Monte Carlo sampling can be performed on the red points, with a cost $\mathcal{O}(\chi^8)$. An exact contraction of the expectation value costs $\mathcal{O}(\chi^{12})$. \label{fig_long_range}}
  \end{centering}
\end{figure}

However, an estimate of the correlator can be obtained using Monte Carlo sampling at a reduced cost (per sample). In \Fig{fig_long_range}, we depict the causal cone structure of two single-site operators separated by $i-j=17$ sites in a binary MERA. The cost of Monte Carlo sampling for $\langle \hat{A}_i \hat{B}_j \rangle$ is just $\mathcal{O}(\chi^{8})$.

This technique can be extended to 2D lattice systems, where calculating long-range correlations exactly quickly becomes infeasible, even for modest values of $\chi$. Note again that memory constraints are particularly challenging for 2D MERA calculations and Monte Carlo sampling can alleviate this burden.

\section{Conclusion and Outlook}

\label{sec_conclusion}

We have outlined and tested a scheme for Monte Carlo sampling with the MERA. Uncorrelated samples can be efficiently generated directly from the wavefunction overlap probability distribution, without needing to resort to Markov chain Monte Carlo methods. From this, expectation values can be extracted and we have demonstrated techniques to reduce the statistical error. We have also presented and demonstrated an algorithm to optimize MERA wavefunctions using sampled energy derivatives.

The numerical results presented here were not intended to be state-of-the-art solutions of the 1D quantum Ising model, but rather to demonstrate feasibility and motivate subsequent applications to 2D systems. In general, Monte Carlo sampling becomes more advantageous for systems with large numbers of degrees of freedom, and we expect 2D MERA to be no exception. Because the reduction in cost in two (and higher) dimensions is so significant, Monte Carlo techniques are a very attractive way to achieve reasonable values of $\chi$ with current computers.

Obvious improvements to the code include utilizing symmetries and parallelization to supercomputers, which is straightfoward with our perfect sampling algorithm. Further research into optimization strategies may lead to other improvements (e.g. by reducing the number of iterations or the tendency to find local minima). Reweighting techniques\cite{Schuch2008} may make the optimization more efficient when approaching the ground state.

\subsection*{Acknowledgements}

The authors would like to thank Philippe Corboz and Anders Sandvik for useful discussions. Support from the Australian Research
Council (FF0668731, DP0878830, DP1092513), the visitor programme at Perimeter Institute, NSERC and FQRNT is acknowledged.

\appendix

\section*{Appendix}

In this appendix we explain how to compute, with cost $O(n^2m)$, the $n\times m$ isometric matrix
\begin{equation}
F = A \exp(A^{\dag} B - B^{\dag}A),
\end{equation}
where $A$ is an $n\times m$ isometric matrix, $B$ is a general $m\times n$ matrix, and $n\le m$. The na\"ive approach would be to evaluate the $m\times m$ matrix $A^{\dag} B - B^{\dag}A$ and compute its exponential, with cost $\mathcal{O}(m^3)$, before multiplying by $A$. However, noting that the exponent does not have full rank (the rank is at most $2n$), we can hope to find a faster method.

Taking the Taylor expansion
\begin{eqnarray}
   F & = &A + B - AB^{\dag}A + \frac{1}{2!}\bigl(BA^{\dag}B - BB^{\dag}A \nonumber \\
   & & - \; AB^{\dag}B + AB^{\dag}AB^{\dag}A \bigr) + \dots,
\end{eqnarray}
we observe that the result can be achieved with a series of multiplications between $n\times n$ matrices $C = A B^{\dag}$, $C^{\dag}$ and $D = B B^{\dag}$, post-multiplied by either $A$ or $B$, requiring total cost $\mathcal{O}(n^2 m)$.

The following algorithm calculates the Taylor expansion to order $p$.
\begin{algorithmic}
\STATE $C \gets A B^{\dag}$; $D \gets B B^{\dag}$
\STATE $W \gets I$; $X \gets 0$; $Y \gets I$; $Z \gets 0$
\FOR{$i = 1 \to p$}
\STATE $\quad\quad T \gets -(WC^{\dag} + XD)/i$
\STATE $\quad\quad X \gets (W + XC)/i$
\STATE $\quad\quad W \gets T$
\STATE $\quad\quad Y \gets Y + W$
\STATE $\quad\quad Z \gets Z + X$
\ENDFOR
\STATE $F \gets YA + ZB$
%
\end{algorithmic}

In the binary MERA, where isometries are $\chi \times \chi^2$ matrices (i.e. $n=\chi$, $m=\chi^2$), the cost of this algorithm scales as $\mathcal{O}(\chi^4)$, compared with the cost $\mathcal{O}(\chi^6)$ of the na\"ive approach. This algorithm becomes particularly important for a tree tensor network and for the MERA in two dimensions, where the na\"ive approach becomes more expensive, in powers of $\chi$, than a sampling (thus becoming the bottle neck of an optimization based on sampling), whereas the above algorithm remains competitive.


\end{document}